\documentclass[article,twocolumn,10pt,nofootinbib]{revtex4-1}% aps,prb,prl, twocolumn,article, book, report,nofootinbib
\usepackage{graphicx,amssymb,amsmath,subfigure,hyperref}%amsart,hyperref
\hypersetup{colorlinks=true, citecolor=blue,linkcolor=red}
\newcommand{\ket}[1]{\left|#1\right>}

\newcommand{\braket}[1]{\left<#1\right>}
\newcommand{\para}[1]{\left(#1\right)}
\newcommand{\abs}[1]{\left|#1\right|}

\begin{document} 
\title{Superconducting analogue of the parafermion fractional quantum Hall states}
\author{Abolhassan Vaezi}
\affiliation{Department of Physics, Cornell University, Ithaca, New York 14853, USA}
\email{vaezi@cornell.edu}

\date{\today}

\begin{abstract}
Read and Rezayi $Z_k$ parafermion wavefunctions describe $\nu=2+\frac{k}{kM+2}$ fractional quantum Hall (FQH) states. These states support non-Abelian excitations from which protected quantum gates can be designed. However, there is no experimental evidence for these non-Abelian anyons to date. In this paper, we study the $\nu=2/k$ FQH-superconductor heterostructure and find the superconducting analogue of the $Z_k$ parafermion FQH state. Our main tool is the mapping of the FQH into coupled one-dimensional (1D) chains each with a pair of counter-propagating modes. We show that by inducing intra-chain pairing and charge preserving backscattering with identical couplings, the 1D chains flow into gapless $Z_{k}$ parafermions when $k< 4$. By studying the effect of inter-chain coupling, we show that every parafermion mode becomes massive except for the two outermost ones. Thus, we achieve a fractional topological superconductor whose chiral edge state is described by a $Z_k$ parafermion conformal field theory. For instance, we find that a $\nu=2/3$ FQH in proximity to a superconductor produces a $Z_3$ parafermion superconducting state. This state is topologically indistinguishable from the non-Abelian part of the $\nu=12/5$ Read-Rezay state. Both of these systems can host Fibonacci anyons capable of performing universal quantum computation through braiding operations.
\end{abstract}

\maketitle

\section{Introduction}

There has been a surge of interest in searching for non-Abelian anyons (non-Abelions)~\cite{RMP-1} in the topological states of matter during the past few years~\cite{RMP-2,RMP-3,Wen-91,MR,PF-1,PF-2,PF-3,FTSC-1}. The non-Abelian states are suitable platforms to perform topological quantum computation via braiding their non-Abelions.
The simplest non-Abelian state is the Pfaffian wavefunction that was proposed by Moore and Read to explain the $\nu=5/2$ fractional quantum Hall (FQH) plateau~\cite{MR}. 
Later, Read and Rezayi generalized Pfaffian to $Z_k$ parafermion wavefunctions as the ground-state of $\nu=2+\frac{k}{kM+2}$ fractional quantum Hall (FQH) states, where $M$ is an even (odd) integer for bosons (fermions)~\cite{RR}. The neutral sector of the edge state can be obtained by computing the correlation functions of an $SU(2)_{k}/U(1)$ conformal field theory (CFT) known as $Z_k$ parafermion CFT~\cite{RMP-1,DF}. These states are believed to have non-Abelian excitations~\cite{RR,RMP-1}. Unfortunately, the experimental search for detecting non-Abelian excitations in the FQH states has failed so far. In this paper, we propose another venue to search for $Z_k$ parafermion states and show that these exotic states emerge in an FQH-superconductor heterostructure. 

The simplest non-Abelian quasiparticle, Ising anyon, was conjectured to be found in the $\nu_c=5/2$ Pfaffian fractional quantum Hall (FQH) state by Moore and Read~\cite{MR}. This state corresponds to $Z_2$ Read-Rezayi parafermions with $M=1$. It was shown later that Majorana fermions can be observed in the superconducting vortices in the weak pairing phase of the $p_x+ip_y$ superconductors as well~\cite{RG,Ivanov} and, these two seemingly different states have identical topological order. This was the first known superconducting state that supports non-Abelions expected in FQH states. 

The most interesting state among the Read-Rezayi wavefunctions is the $Z_3$ parafermion state that describes the ground-state of $\nu=12/5$ and $\nu=13/5$ FQH states~\cite{RR,RMP-1}. This state state supports Fibonacci anyon, a non-Abelion that has $d_{F}=\frac{1+\sqrt{5}}{2}$ quantum dimension and can perform universal quantum computation, i.e., all the quantum gates can be designed and measured by braiding Fibonacci anyons~\cite{RMP-1,Fib-1,Fib-2,Fib-3}. The quantum dimension of Fibonacci anyons can be understood by counting the number of degenerate ground-states at the presence of $n$ Fibonacci anyon excitations. For that purpose  we need to consider their fusion algebra: $\tau \times \tau \sim 1+\tau$, which means a state with $n$ Fibonacci anyons can be mapped into the superposition of states with two and one less Fibonacci anyons, respectively. Thus, the ground-state degeneracy (GSD) in the presence of $n$ Fibonacci anyon excitations, $G\para{n}$, satisfies the Fibonacci recursion relation, i.e., $G\para{n}=G\para{n-1}+G\para{n-2}$. At the large $n$ limit, GSD grows as $\log G\para{n}\sim n\log d_{F}+...~$.

Achieving the two-dimensional (2D) superconducting analogue of the Read-Rezayi parafermion states requires answering two major steps: First, in what system can we observe parafermions~\cite{Fradkin,Duality-1,PF-0,Cobanera} as the physical degree of freedom? Second, how can we condense such quasi-particles? Parafermions are the generalization of the Majorana fermion, sometimes called fractionalized Majorana fermions. Consider $\chi_{i}$ parafermion operator at site $i$.  A $Z_k$ parafermion operator satisfies $\chi_{i}^{k}=1$, $\chi_{i}^{\dag}=\chi_{i}^{k-1}$, and $\chi_{i}\chi_{j}=\exp\para{2\pi i/k}\chi_{j}\chi_{i}$ algebra for $i<j$~\cite{Fradkin,PF-0}.  The $Z_k$ parafermion Read-Rezayi state can be obtained by condensing a cluster of $k$ parafermions. This condensation is allowed because $k$ parafermion cluster behaves like a bosonic object. Fibonacci anyons emerge as the excitations above the condensate of $Z_3$ parafermions. For instance, they can bind to the topological defects of the condensate, e.g., inside vortices.

Recently, the problem of perturbing a counter-propagating edge mode of a fractional topological insulator~\cite{FTI-1} or two nearby FQH states with opposite spin polarizations by either electron pairing or backscattering has attracted a lot of attention. It has been shown that the parafermion (fractionalized Majorana fermion) zero modes can be obtained at the domain wall between regions with these different mass terms~\cite{PF-1,PF-2,PF-3,Beigi,PF-5,PF-6,PF-7,Braid-1,Braid-2}. It has also been shown that parafermion zero modes can be found in the bulk of an FQH state which has acquired superconducting pairing through proximity effect~\cite{FTSC-1} as well as in other 2D systems~\cite{Bombin,Genon-0,Genon-1,PF-4,Genon-4,Genon-2,Genon-3,Genon-5}.

\begin{figure}[t]\label{fig1}
\includegraphics[width=.35 \textwidth]{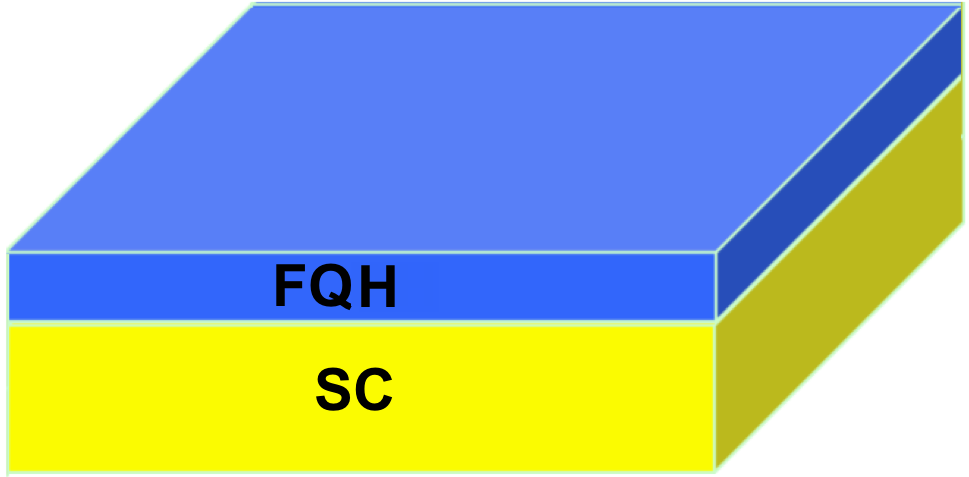}
\caption{Schematic setup of a fractional topological superconductor. This state of matter can be achieved by inducing superconducting order parameter into a fractional quantum Hall bar or a fractional Chern insulator through the proximity effect.
} 
\end{figure}

After obtaining parafermions, a condensate of parafermions is needed to achieve a $Z_k$ parafermion state. In this paper we show that this can naturally happen in a fractional topological superconductor (FTSC) which is defined as an Abelian FQH state that has acquired superconducting pairing either intrinsically or through proximity effect (see Fig. \ref{fig1}). The basic observation behind our idea is the way we obtain $Z_k$ Parafermion FQH states. It is achieved by condensing clusters of $k$ quasiparticles. Now imagine an Abelian FQH state at $\nu=2/k$. It supports anyon excitations with $q_{a}=2e/k$ electric charge. We then induce pairing by some means into this state. The charge of the Cooper pair is $2e$, therefore, we need {\em $k$-anyon condensation}. This condensate  is topologically indistinguishable from the non-Abelian part of the Read-Rezayi $Z_k$ parafermion wave-function. %$k$-anyon condensation allows us to consider $f_i^\dag \propto\para{ a_i^\dag+a_i^{k-1}}$ operator, where $a_i$ is the anyon operator with $\theta=2\pi/k$ exchange statistics, as a physical local excitation. It is easy to check that $f_i$ satisfies $Z_k$ parafermion algebra. 

The Fractional Chern insulator (FCI) is a topological state of strongly interacting electrons residing in a partially filled nearly flat band with nontrivial Chern number~\cite{FCI-1,FCI-2,FCI-3,FCI-4,FCI-5,FCI-6,FCI-7}. FCI has the same topological order as an FQH though there is no external magnetic field. Therefore, as long as the topological order of the parent state is concerned, we can substitute FQH for FCI without changing the final results. From now on, we do not distinguish between them and one can replace FQH with FCI everywhere in this paper. It is worth mentioning that there are two huge advantages for the FCI compared to the FQH in our discussions. First, there is no external magnetic field that can kill superconductivity in an FQH system. Second, it is much easier to induce s-wave pairing in it. 

These intriguing findings motivated us to reexamine the superconductor (SC)-FQH heterostructure more carefully. However, the nature of this problem is difficult and highly nontrivial due to strong correlations in the FQH state. Nevertheless, we introduce a trick that can help us understand this situation better.  We will discuss in section \ref{sec:two} that a generic Abelian FQH state can be mapped into an array of one dimensional (1D) chains of counter-propagating free bosons. These chains are coupled by anyon back-scattering terms such that bulk becomes fully gapped while boundaries host chiral gapless modes. A similar construction has been introduced by Teo and Kane to obtain FQH states by starting from decoupled Luttinger liquids. This type of mapping helps us to study the effect of superconducting pairing on these 1D chains more efficiently by utilizing powerful techniques of CFT approach. CFT predicts that when $(\nu_c$ mod $1) < 1/2$, the 1D theory of each chains flows to $Z_{k}$ parafermions in the infra-red (IR) fixed point~\cite{Lech-1,Lech-2}, where $k=2/{\nu}$. Inter-chain anyon backscattering gaps out the bulk of the system but maintains a chiral edge state described by $Z_{k}$ parafermion CFT around the sample.

It is worth mentioning that the FQH state has a non-zero bulk gap that prevents quantum phase transition as long as it is open. When the superconducting pairing is weak the bulk gap does not close and no phase transition is expected. As the pairing becomes stronger and comparable to the many body gap, we reach a quantum critical point. The system undergoes a topological phase transition into a different topological state at this point and when gap reopens, the nature of the state becomes completely different than its parent state. This type of problem has been first studied by us in Ref.~[\onlinecite{FTSC-1}]. We showed there that the high energy description of the FTSC can be studied by modifying the edge theory of the Abelian parent state. Assuming the parent state's edge theory is $U(1)_{k}$ CFT,  after the phase transition it would be a $U(1)_{k}/Z_2$ orbifold CFT~\cite{Dijkgraaf,DF,BW-0, BW-1,FTSC-1}. In this paper, we reinvestigate this problem and show that for $k<4$, the system has a different low energy description than its high energy limit. For these values of $k$, we can find the superconducting analogue of the Read-Rezayi $Z_{k}$ parafermion state.   

Our paper is organized as follows. 
In Section \ref{sec:two}, we briefly introduce the 2D bosonization of the Abelian FQH states by mapping them to an array of 1D chains through applying a similar construction to that of Teo and Kane~\cite{Teo-Kane}. This mapping allows us to use powerful techniques of the CFT approach. Using the bosonization framework, we study the FQH-SC heterojunction and map this problem to coupled 1D chains as well. In section \ref{sec:three}, we focus on the Hamiltonian that describes a single chain. We discuss that the self-dual sine-Gordon (SDSG) model emerges as the effective Hamiltonian of the decoupled 1D chains. We show that by tuning the strength of inter-chain pairing and backscattering we can achieve the self-dual point of the SDSG model which is a quantum critical point. It is believed that this critical point is described by a $Z_{k}$ parafermion CFT for $k<4$. One mechanism that can enhance backscattering is to induce in plane magentic order in an FCI with strong spin-orbit coupling. This can be achieved by putting a ferromagnet (FM) on top of an FCI. In section \ref{sec:four}, we study the effect of inter-chain coupling and show that it gaps out every two nearby counter-propagating modes. This observations solves our main problem of finding the superconducting analogue of the Read-Rezayi  $Z_k$ parafermion FQH state.

\section{Bosonization of Abelian FQH states and FTSCs}\label{sec:two}

As we mentioned earlier in the current paper, an FCI is smoothly connected to the FQH state in presence of external magnetic field. Therefore, they are described by the same topological order, i.e., their  anyon excitations, edge states, and every other topological characteristic, e.g., the ground-state degeneracy is identical. In this section, FQH at filling $\nu_c$ denotes all the gapped physical systems that can be smoothly transformed into one another, including FCI.

The FQH effect requires the interaction between electrons to be much larger than their kinetic energies. Therefore, the FQH state is by definition a strongly correlated system and any description of the system is approximate and far from being exact. Thus, the problem of FQH-SC heterostructure is very challenging. In this section we pause and try to overcome this difficulty first. We introduce a mapping that will turn out to be very useful in solving our problem. We demonstrate that an Abelian FQH state can be mapped into an array of coupled 1D chains of counter-propagating gapless modes each described by a free boson CFT. This mapping to coupled 1D chains allows us to take advantage of the powerful techniques of the CFT approach in solving our main problem.

\begin{figure}[t]\label{fig2}
\includegraphics[width=.5 \textwidth]{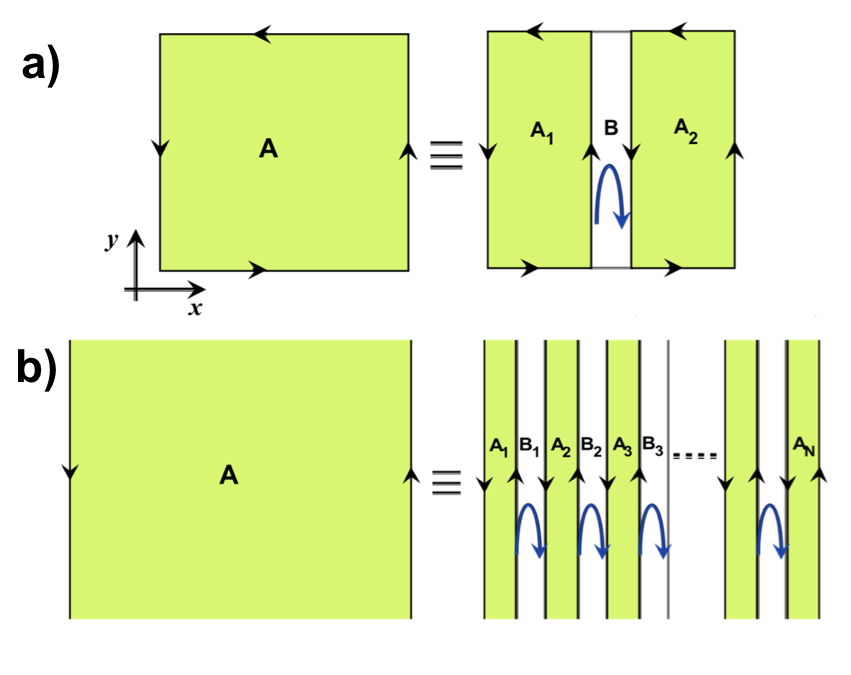}
\caption{An Abelian FQH state can be imagined as an array of coupled one dimensional chains each described by a free boson CFT.  a) FQH region $A$ can be split into regions $A_1$ and $A_2$ in addition to the extremely narrow vacuum region $B$. Bent blue arrow represents backscattering process between the two counter-propagating gapless modes.  b) By repeating this procedure we can map a 2D FQH state into an array of narrow strips $A_1$ to $A_{\mathcal{N}}$ that are coupled to one another through backscattering. Each narrow strip can be imagined as a 1D chain with a free boson description.
} 
\end{figure}

The basic observation behind our scheme is that the interior of an Abelian FQH state is gapped while its boundary with a trivial insulator, e.g.,  vacuum supports gapless edge modes described by a free boson CFT~\cite{Wen-Book}.  Furthermore, we assume that every two systems with gapped bulk and the same edge theory are described by the same topological order. Next, we cut the FQH system into three pieces: $A_1$ and $A_2$, that are separated by a third trivial narrow region $B$ (see Fig. \ref{fig2}) whose length along the $x$ axis will be set to zero at the end of the process. Since both $A_1$ and $A_2$ regions are surrounded by vacuum, each support a gapless chiral edge mode. Therefore there are two counter-propagating modes on the two sides of region $B$. It should be emphasized that these two counter-propagating modes can have identical center of mass momenta. Because we take the narrow limit of the intervening region, the two counter-propagating modes can couple to one another through backscattering. It is well-known that the backscattering between two identical counter-propagating modes can gap them out~\cite{Lu-1,Juven-1}. Therefore, there is no gapless edge state on the two sides of region $B$ after taking the backscattering process into consideration. This is consistent with our intuition of an FQH state that can have gapless edge state only at its boundary with a different topological state.

The above procedure can be easily generalized by splitting the system into $\mathcal{N}$ FQH regions, $A_{I}$, that are separated by narrow vacuum regions $B_{I}$ (see Fig. \ref{fig2}). After taking the $\mathcal{N}\to \infty$ limit, the narrow FQH liquid region $A_{I}$ becomes a 1D chain with two counter-propagating modes. These chains are coupled by backscattering through $B_{I}$ regions. This simple trick motivates us to imagine an FQH state as an array of coupled 1D chains, thus suggesting a CFT description of the system. To summarize, we first introduce an infinite number of narrow strips each with two counter-propagating modes. The strips acquire mass through backscattering between adjacent counter-propagating modes. In this way the bulk of the system becomes gapped. If the two outermost modes at the boundary of the original sample remain gapless the system corresponds to an FQH, otherwise it represents a trivial insulator with trivial topological order. In the rest of this paper we demonstrate how powerful this technique is in understanding the superconducting proximity effect on the Abelian FQH states.

\subsection{Bosonization of fractional quantum states}

Consider an Abelian FQH state of electrons at filling fraction $\abs{\nu_c}=2/k$ where $N\in \mathbb{Z}$. For $k=2m$ it can be a monolayer, while for $k=2m+1$ we consider a bilayer system, e.g., $(k,k,0)$, $(k-1,k-1,1)$ or $(1,1,k-1)$ Halperin states. The multi-component quantum Hall states can have extended symmetries. Hence, excitations carry different charges for example pseudo-spin (associated with layer index), spin, and electric charge. From now on, for two reasons we focus on the charged sector of the system and charged degrees of freedom only. Firstly, neutral fluctuations are unaffected by inducing superconducting pairing and do not modify our results. Secondly, we need to somehow gap out the neutral fluctuations for a simple reason. The minimum charge of excitations in the Halperin states mentioned earlier is $\nu_c/2=1/k$ rather than $\nu_c=2/k$. However, we need {\em $k-$anyon condensation} in order to obtain a $Z_k$ symmetric state, hence the minimum charge of excitation should become twice bigger. This can be achieved through binding anyons of the top and bottom layers, i.e., gapping the pseudo-spin excitation. 

Before discussing the bosonization procedure let us pause and briefly describe how we can decouple charge and neutral degrees of freedom in the $(nnl)$ two-component quantum Hall state with $\nu_c=\frac{2}{n+l}$ filling fraction. Using the $K$ matrix formulation, there are $\det(K)=n^2-l^2$ non-trivial anyon excitations described by integer valued $\vec{p}\equiv (p_1,p_2)$ vectors subject to $\vec{p}\sim \vec{p}+(n,l)$, $\vec{p}\sim \vec{p}+(l,n)$ identifications. The self-statistics of $\vec{p}$ excitation is $\theta_{\bf p,p}=\pi \bf p^{\rm T} \bf K^{-1} p$ and the mutual statistics between $\vec{p}$ and $\vec{q}$ distinct excitations is $\theta_{\bf p,q}=2\pi \bf p^{\rm T} \bf K^{-1} q$~\cite{Wen-Book}.  There are two electron operators given by $(n,l)$ and $(l,n)$ vectors. In the edge CFT picture, the electron operators can be bosonized as $c_{+}=e^{i\para{n\phi_1+l\phi_2}}$ and $c_{-}=e^{i\para{l\phi_1+n\phi_2}}$ with $\left\langle\phi_i(z)\phi_j(w)\right\rangle= \para{K^{-1}}_{ij} \log \para{z-w}$ correlation function. For the $(nnl)$ state we can define the charged and neutral chiral bosons as $\phi_{c}=\para{\phi_{1}+\phi_{2}}$, and $\phi_{s}=\para{\phi_{1}-\phi_{2}}$. Thus, the two electron operators are represented by $c_{+}=e^{i\para{\phi_c/\nu_c+\phi_s\nu_s}}$ and $c_{-}=e^{i\para{\phi_c/\nu_c-\phi_s/\nu_s}}$ in terms of charged and neutral degrees of freedom where $\nu_c=\frac{2}{n+l}$ and $\nu_s=\frac{2}{n-l}$. Later we imagine a situation where the neutral sector is gapped. This assumption allows us to drop the neutral boson from expressions and both electron operaters can be represented as $c \propto e^{i\phi_c/\nu_c}$ where $\nu_c$ is the filling fraction. 

Now let us come back to our coupled wires construction of the FQH states. The procedure explained in the previous section suggests the following description for the ``charged sector" of an Abelian FQH state at filling fraction $\nu_c$ in terms of coupled 1D chains~\cite{comment-1,PF-2}
\begin{eqnarray}\label{eq:bos-1}
&&H_0=\int dx~\para{\mathcal{H}_0+\mathcal{H}^{BS}_1+\mathcal{H}^{BS}_2},\cr
&&\mathcal{H}_0=~~~~~\sum_{I=1}^{{\mathcal{N}}}\sum_{\tau=c,s}~\frac{\nu_{\tau}^{-1}}{4\pi} \left[\para{\partial_{x}\phi^{I}_{\tau R}}^2+\para{\partial_{x}\phi^{I}_{\tau L}}^2\right],\cr
&&\mathcal{H}^{BS}_1=-\sum_{I=1}^{{\mathcal{N}}}~g^{BS}_{I,I}\cos\para{\frac{\phi^{I}_{cR}-\phi^{I}_{cL}}{\nu_c}}\cos\para{\frac{\phi^{I}_{sR}-\phi^{I}_{sL}}{\nu_s}},\cr
&&\mathcal{H}^{BS}_2=-\sum_{I=1}^{{\mathcal{N}-1}}~g^{BS}_{I,I+1} \cos\para{\frac{\phi^{I}_{cR}-\phi^{I+1}_{cL}}{\nu_c}}\cos\para{\frac{\phi^{I}_{sR}-\phi^{I+1}_{sL}}{\nu_s}},\cr
&&
\end{eqnarray}
where $\exp\para{i\phi^{cI}_{R/L}/\nu_c\pm \phi^{sI}_{R/L}/\nu_s}$ denotes the bosonized right/left moving electron operator on the upper/lower layer and the $I$-th chain. Here $\mathcal{H}^{BS}_{1}$ represents the intra-chain backscattering of electrons and $\mathcal{H}^{BS}_{2}$ inter-chain electron backscattering. In principle the ``{\em intra-chain anyon backscattering"} process is allowed and must be taken into account in our analysis. However, as we will discuss shortly, for the setup we study in this paper they are subdominant terms and we will neglect them in our discussion. On the other hand, the inter-chain anyon backscattering is strictly disallowed since anyons cannot pass through the vacuum (see Fig. \ref{fig2}). 

In the quantum Hall regime, $\mathcal{H}^{BS}_{1}$ as well as intra-chain anyon backscattering terms should be either absent or weak enough, otherwise all chains become gapped and there will be no edge state. However, there are ways to generate {\em intra-chain electron backscattering} ($\mathcal{H}^{BS}_{1}$). For example, inducing an in-plane magnetic order in the fractional Chern insulator with strong spin-orbit coupling through proximity to a ferro-magnet can enhance electron backscattering.  The presence of strong spin-orbit coupling leads to (partial) spin momentum locking, hence spin flip will amount to backscattering of electrons. This effect does not enhance anyon backscattering. The reason is that anyon backscattering through proximity effect requires anyon excitation in the substrate as well whereas ferro-magnet is a topologically trivial state and does not support anyon excitations. {\em Therefore, in this paper we make the central assumption that proximity induced intra-chain electron backscattering is much stronger than intrinsic intra-chain anyon backscattering because proximity effect can enhance electron backscattering only}. The above simple model in Eq. \eqref{eq:bos-1} can describe the phase transition between the FQH and the trivial state. For instance, when $g_{I,I}\ll g_{I,I+1}$ we can ignore $\mathcal{H}^{BS}_{1}$, so $\phi^{1}_{L}$ and $\phi^{N}_{R}$ which live at the boundaries of the system remain gapless, while all other modes become massive. Therefore, we obtain an FQH state for $g_{I,I}\ll g_{I,I+1}$. In the opposite limit where $g_{I,I}\gg g_{I,I+1}$ we can ignore $\mathcal{H}^{BS}_{2}$ and due to strong intra-chain backscattering the system is completely gapped, resulting in a trivial insulator. 

\subsection{Bosonization of fractional topological superconductors} 

We now wish to study the effect of superconductivity on an Abelian FCI state at filling fraction $\nu_c=2/k$. To this end, we assume the spin orbit coupling is strong in the system, and thus the electron spin is correlated with its momentum. In other words, the spin of $c^{I}_{\pm,R}\sim e^{i\phi^{I}_{cR}/{\nu_c}\pm i\phi^{I}_{sR}/{\nu_s}}$  electron operator is opposite to that of $c^{I}_{\pm,L}\sim e^{i\phi^{I}_{L}/{\nu_c\pm i\phi^{I}_{sL}/{\nu_s}}}$. Consequently, they can be paired through s-wave pairing. This s-wave superconductivity can be either intrinsic or induced through the proximity effect. We consider two perturbations: intra-chain and inter-chain pairing which have $g^{SC}_{I,I} c^{I}_{R}c^{I}_{L}$ and $g^{SC}_{I,I+1} c^{I}_{R}c^{I+1}_{L}$ structures, respectively. These terms have the following bosonized representations,
\begin{eqnarray}
&&\mathcal{H}^{SC}_1=-\sum_{I=1}^{{\mathcal{N}}}~g^{SC}_{I,I} \cos\para{\frac{\phi^{I}_{cR}+\phi^{I}_{cL}}{\nu_c}}\cos\para{\frac{\phi^{I}_{sR}-\phi^{I}_{sL}}{\nu_s}},\cr
&&\mathcal{H}^{SC}_2=-\sum_{I=1}^{{\mathcal{N}-1}}~g^{SC}_{I,I+1} \cos\para{\frac{\phi^{I}_{cR}+\phi^{I+1}_{cL}}{\nu_c}}\cos\para{\frac{\phi^{I}_{sR}-\phi^{I+1}_{sL}}{\nu_s}}.\cr
&&
\end{eqnarray}
Therefore, the Hamiltonian density of an FTSC is $\mathcal{H}_0+\mathcal{H}^{BS}_1+\mathcal{H}^{BS}_2+\mathcal{H}^{SC}_1+\mathcal{H}^{SC}_2$. Now let us define the following bosonic variables: 
\begin{eqnarray}
&&\varphi^{I}_{c/s}=\frac{\phi^{I}_{c/s,R}+\phi^{I}_{c/s,L}}{2},\quad \theta^{I}_{c/s}=\frac{\phi^{I}_{c/s,R}-\phi^{I}_{c/s,L}}{2}.
\end{eqnarray}
with
\begin{eqnarray}
&&\left[\varphi^{I}_{\tau}\para{x,t},\varphi^{I}_{\tau'}\para{y,t}\right]=0,\quad \left[\theta^{I}_{\tau}\para{x,t},\theta^{I}_{\tau'}\para{y,t}\right]=0,\cr
&&\left[\varphi^{I}_{\tau}\para{x,t},\theta^{I}_{\tau'}\para{y,t}\right]=i\frac{\pi\nu_{\tau}}{2}\delta_{\tau,\tau'}{\rm sgn}\para{x-y},
\end{eqnarray}
equal time commutation relations. On one hand, the above bosonic variables tend to fluctuate due to the Kinetic term $\mathcal{H}_0$. On the other hand, in order to gain potential energy the argument of the cosine perturbations must freeze, i.e., replaced by a constant value (c-number) at every spatial point. If the latter happens, then the sector associated with that bosonic variable becomes massive. However, this wish cannot be always fulfilled. For example, if we condense $\varphi^I_c\para{x,t}$ boson (i.e., replace it by its expectation value) such that $\abs{\cos\para{2\varphi^I_c/\nu_c}}=1$, then we cannot have $\abs{\cos\para{2\theta^I_c/\nu_c}}=1$, because  $\left[\varphi^{I}_{c}\para{x,t},\theta^{I}_{c}\para{y,t}\right]\neq 0$, and these two conjugate variables cannot simultaneously condense. An important observation about the above Hamiltonians is that it contains $\cos\para{2\theta^I_s/\nu_s}$ only and as a result we can maximize it by assuming $\theta^I_s=n\nu_s\pi$, where $n$ is an integer number. Thus, the neutral sector is gapped out. For this reason, from now on we focus on the charged sector and investigate the effect of having two incompatible perturbations $\cos\para{2\varphi^I_c/\nu_c}$ and $\cos\para{2\theta^I_c/\nu_c}$. % bosonic variable commutes with all other terms in the Hamiltonian, as well as itself at a different spatial point, thus it is a constant of motion and can be replaced by its classical value that minimizes $\cos\para{\frac{\phi^{I}_{sR}-\phi^{I}_{sL}}{\nu_s}}$. However, neither $\para{\phi^{I}_{cR}-\phi^{I}_{cL}}$ nor $\para{\phi^{I}_{cR}+\phi^{I}_{cL}}$  can be condensed because these are noncommuting variables and  the Hamiltonain contains both of them.}

In the absence of inter-chain interaction and after condensing $\theta^I_s$ neutral bosons, the above Hamiltonian reduces to an array of ${\mathcal{N}}\to \infty$ decoupled 1D chains each described by a SDSG Hamiltonian~\cite{Lech-1}:
\begin{eqnarray}\label{SG-1}
\mathcal{H}^{I}_{SG}=\frac{\nu_c^{-1}}{2\pi}&&\int dx~ \left[\para{\partial_{x}\varphi^{I}_c}^2+\para{\partial_{x}\theta^{I}_c}^2\right],\cr
-&&\int dx ~\left[g^{*~SC}_{I,I}\cos\para{2\varphi_c/\nu_c}+g^{*~BS}_{I,I}\cos\para{2\theta_c/\nu_c}\right].~~~
\end{eqnarray}
where $g^{*~SC/BS}_{I,I}=g^{SC/BS}_{I,I}\left\langle\cos\para{2\theta^I_s/\nu_s}\right\rangle$.
%\begin{eqnarray}
%&&\varphi^{I}=\frac{\phi^{I}_{R}+\phi^{I}_{L}}{2},\quad \theta^{I}=\frac{\phi^{I}_{R}-\phi^{I}_{L}}{2}.
%\end{eqnarray}
In the following section, we first try to review the properties of the above well-studied SDSG model~\cite{Lech-1,Lech-2}. We then investigate the effect of coupling between different 1D chains that are described by the above SDSG Hamiltonian. This will help us to find a route for finding the superconducting analogue of the Read-Rezayi parafermion wavefunction.
 
\section{Phase diagram of decoupled 1D chains}\label{sec:three}

In this section we focus on one chain of two counter-propagating modes perturbed by superconducting pairing and electron back-scattering. As we mentioned in the previous section, the resulting Hamiltonian has the SDSG form. So, let us consider the following SDSG Hamiltonian of a 1D system:
\begin{eqnarray}\label{SG-2}
\mathcal{H}_{SG}\para{k}=\frac{k}{4\pi}&&\int dx~ \left[\para{\partial_{x}\varphi}^2+\para{\partial_{x}\theta}^2\right],\cr
-&&\int dx ~\left[g_{1}\cos\para{k\varphi}+g_{2}\cos\para{k\theta}\right].
\end{eqnarray}
By comparing the above equation with Eq. \eqref{SG-1}, we see that $k$ is related to the filling fraction of the parent FQH state as follows: $k=2/\nu_c$. The first part of the SDSG Hamiltonian describes a free boson theory with $c=1$ compactified on a circle with radius $R=\sqrt{k/2}$. The second part of the SDSG Hamiltonian contains two different types of mass terms that do not commute with each other, since:     
\begin{eqnarray}
\left[\varphi\para{x},\theta\para{x'}\right]=i\frac{\pi}{k}{\rm sgn}\para{x-x'},
\end{eqnarray} 
The conformal dimension of the $g_1$ as well as $g_2$ term is $k/2$. Therefore, they are relevant perturbations when $k< 4$. The cosine perturbations break the $U(1)\times U(1)$ symmetry associated with the kinetic part of the SDSG model to its $Z_k \times Z^{\rm dual}_k$ subgroup generated by 
\begin{eqnarray}\label{Sym-1}
&&Z_k:\quad \varphi \to \varphi+2\pi/k,\quad \theta \to \theta,
\end{eqnarray}
and 
\begin{eqnarray}
&&Z^{\rm dual}_k:\quad \theta \to \theta +2\pi/k, \quad \varphi \to \varphi,
\end{eqnarray}
respectively. Interestingly, the $Z_k$ clock model also enjoys two distinct $Z_k$ symmetries. In fact, it can be shown that the SDSG model with index $k$ describes the continuum limit of the $Z_k$ clock model~\cite{clock-3}.

Some useful insights can be obtained by studying the SDSG model at its limiting cases. For instance,  for $g_2=0$, the charged $\varphi$ condenses at one of its classical values, i.e., $\varphi_0=\frac{2\pi m}{k}$. This can happen since $\left[\varphi\para{x},\varphi\para{x'}\right]=0$. Similarly, when $g_1=0$, $\theta$ field condenses to its classical values $\theta_0=\frac{2\pi m}{k}$, because $\left[\theta\para{x},\theta\para{x'}\right]=0$. Let us define the following order parameters 

\begin{eqnarray}\label{order-1}
&&\sigma\para{z,\bar{z}} \sim e^{i\varphi\para{z,\bar{z}}}+\alpha e^{-i\para{k-1}\varphi\para{z,\bar{z}}},\cr
&&\mu\para{z,\bar{z}}\sim e^{i\theta\para{z,\bar{z}}} +\beta e^{-i\para{k-1}\theta\para{z,\bar{z}}},
\end{eqnarray}
where $\alpha$, and  $\beta$ are two non-universal constant. The $\sigma$ order parameters (also known as spin field) carries unit charge of the first $Z_k$ and neutral under the dual $Z_k$ symmetry, while $\mu$ order parameter (also known as spin disorder field) is neutral under $Z_k$  and carries unit charge of the $Z_k^{\rm dual}$. These two order parameters do not commute and cannot condense simultaneously. Therefore, we can distinguish two gapped phases by looking at the expectation values of $\sigma$ and $\mu$ fields. The so-called ferromagnetic phase ($g_1\gg g_2$) is characterized by $\braket{\sigma}\neq 0$, and $\braket{\mu}=0$, while in the paramagnetic phase ($g_1\ll g_2$) by $\braket{\mu}\neq 0$, and $\braket{\sigma}=0$.

The SDSG model at $g_1 g_2=0$ reduces to the $XY$ model of a system with in plane magnetic order that has a single global $U(1)$ symmetry associated with either $\varphi$ or $\theta$ field, and a discrete $Z_k$ symmetry associated with the other one. On the other hand, the $g_2$ ($g_1$) mass term can be imagined as the vortex operator for the $g_1$ ($g_2$) mass term. This vortex operator breaks the mentioned $U(1)$ symmetry down to its $Z_k$ subgroup. To see this we need to consider the following commutation relation
\begin{eqnarray}
&&\left[\varphi\para{x},e^{-ik\theta\para{x'}}\right]=\pi {\rm sgn}\para{x-x'} e^{-ik\theta\para{x'}}.
\end{eqnarray} 
As a result, if we label a ground-state by $\varphi\para{x}\ket{\varphi_0}=\varphi_0\ket{\varphi_0}$, we have
\begin{eqnarray}
&&e^{-ik\theta\para{x'}}\ket{\varphi_0}\sim \ket{\varphi_0+\pi}\quad x'<x,\cr
&&e^{-ik\theta\para{x'}}\ket{\varphi_0}\sim \ket{\varphi_0-\pi }\quad x'>x,
\end{eqnarray}
which simply means $e^{-ik\theta\para{x'}}$ creates a kink in the profile of $\varphi_0$ at $x'$ and shifts its value by $2\pi$ at that point. Consequently, $g_2$ term creates vortices and compactifies $\varphi$ field as
\begin{eqnarray}
\varphi \sim \varphi +2\pi.
\end{eqnarray}
Due to this periodicity, $\varphi_0$ has only $k$ distinct values given by
\begin{eqnarray}
\varphi_0=\frac{2\pi m}{k}, \quad m=0,...,k-1.
\end{eqnarray}

Therefore, when $g_1\gg g_2$, $\varphi$ condenses and the free boson theory becomes massive. Similarly, when $g_2\gg g_1$, $\theta$ condenses and again we obtain a gapped theory. However, the situation is fundamentally different at the $\abs{g_1}=\abs{g_2}$ point. In this case, the theory is symmetric under $\varphi \leftrightarrow \theta$ exchange and shows self-duality. At this point neither $\varphi$ nor $\theta$ can condense and is a  critical point~\cite{Lech-1,Lech-2,Teo-Kane}.

At the ultra-violet (UV) fixed point the SDSG model is a CFT with ${\rm c_{UV}=1}$. At the infra-red (IR) limit, the SDSG model is critical at the self-dual point and described by a CFT. On the other hand, according to the c-theorem~\cite{Zamolodchikov} when we perturb a CFT  with relevant operators (in the sense of RG) i.e., when $k< 4$ in our problem (so the scaling dimension of the mass terms does not exceed two) then ${\rm c_{IR} < c_{UV}=1}$. Hence, the IR fixed point is necessarily described by a minimal CFT.  At this IR fixed point, $g_1$ and $g_2$ terms compete with each other and do not let the other term to condense. This can be seen through the vortex picture. We can approach the self-dual point from the $g_1>g_2$ side. Along that direction $g_1$ term dominates over the $g_2$ term, and thus $\cos\para{k\varphi}$ maximizes through picking  one of $k$ distinct values of $\varphi$. Since $\theta$ and $\varphi$ are conjugate fields, $g_2\cos\para{k\theta}$ term creates vortices for the classical configuration of $\varphi_0\para{x}$. Therefore, $g_2$ can be considered as the fugacity of the vortex gas. When $g_2$ approaches $g_1$, vortices proliferate all over the system and $\varphi$ fluctuates strongly. Thus at $g_2=g_1$, a second order phase transition of the BKT type (vortex proliferation) is expected. Since, $\varphi_0$ can take $k$ distinct values, the universality class of the phase transition is identical to that of the $Z_k$ clock model or equivalently to that of a $k$-state Potts model. The critical point of these theories is described by a $Z_k$ parafermion CFT~\cite{Lech-1}.  Therefore, we expect that for $k<4$ the SDSG model to be described by a $Z_k$ parafermion CFT at its IR fixed point. 

When $k>4$, both $g_1$ and $g_2$ terms are irrelevant, therefore, $c_{UV}=c_{IR}$. It is known that for $u<u_{c1}<1$, where $u=g_2/g_1$, $\varphi$ condenses, while for $u>u_{c2}>1$ the conjugate field $\theta$ is condensed. For these values of $k$, instead of having a single critical point, the system is gapless with Gaussian fluctuations all over the so called Villain line: $u_{c1}<u<u_{c2}$~\cite{clock-1,clock-2}. 
Furthermore, Fendley showed that after a generalized Jordan-Wigner
transformation, the $Z_k$ chiral quantum clock model transforms to a dual theory of
a $Z_k$ parafermion chain, which exhibits an end state $d=\sqrt{k}$ quantum dimension when the original
clock model is in its ordered phase~\cite{PF-0}. So, we conclude that the CFT description of the SDSG must allow for a primary field with the same quantum dimension.

%Furthermore, Fendley has shown that 1D open chain of the $k$-state chiral quantum clock model supports $Z_k$ parafermion end state with $d=\sqrt{k}$ quantum dimension~\cite{PF-0}. So, we conclude that the CFT description of the SDSG at its UV fixed point must allow for a primary field with the same quantum dimension.

For $k< 4$, the UV fixed point of the SDSG model i.e., the unperturbed Hamiltonian (in which $g_1=g_2=0$) is described by $U(1)_{k}$ free boson CFT whose central charge is $c=1$. Since, the perturbations in the SDSG model is relevant for $k< 4$, the IR fixed point of the SDSG model has a different CFT description. For example, for $k=2$ we achieve a single Ising CFT ($Z_2$ parafermion) and a $Z_3$ parafermion CFT for $k=3$ in the IR limit. Furthermore, noting the facts that $U(1)_k/Z_2$ orbifold CFT has $Z_k$ symmetry, $c=1$ central charge, and twist fields with quantum dimension $d_{tw}=\sqrt{k}$ that trap $Z_k$ parafermion zero mode~\cite{FTSC-1}, we conjecture that the gapless region of the SDSG for $k>4$ is described by the $U(1)_{k}/Z_2$ orbifold theory.

%For $k< 4$, it can be shown that UV fixed point of the SDSG model (which is related to the clock model) is described by $U(1)_{k}/Z_2$ orbifold CFT whose central charge is $c=1$. The $U(1)_k/Z_2$ theory is equivalent to the tensor product of two decoupled Ising CFTs, $Z_4$ parafermion theory, and the critical point of four-state Potts for $k=2,3,4$ cases respectively~\cite{Dijkgraaf}. Since, the perturbations in the SDSG model is relevant for $k< 4$, for $k=2$ one of the Ising sectors in the $U(1)_2/Z_2$ orbifold CFT becomes massive and we reach a single Ising CFT ($Z_2$ parafermion) in the IR limit. Similarly, the orbifold theory flows to $Z_3$ parafermion CFT for $k=3$. Noting the facts that $U(1)_k/Z_2$ orbifold CFT has $Z_k$ symmetry, $c=1$ central charge, and twist fields with quantum dimension $d_{tw}=\sqrt{k}$ that trap $Z_k$ parafermion zero mode~\cite{FTSC-1}, we conjecture that the gapless region of the SDSG at $k>4$ is also described by $U(1)_{k}/Z_2$ orbifold theory.

Before closing this section, it is worth mentioning that ``{\em parafermion primary field}" $\psi_{1,R}$ of $Z_k$ parafermion CFT carries unit charge of both $Z_k$ and $Z_k^{\rm dual}$ symmetries in Eq. \eqref{Sym-1}. Similarly,  the $\psi_{1,L}$ carries +1 $Z_k$ charge and -1 $Z_k^{\rm dual}$ charge~\cite{clock-3}. This observation suggests the following identifications in terms of the UV fixed point free boson fields:
\begin{eqnarray}\label{Parafermion-1}
&&\psi_{1,R}\sim \sigma\para{z,\bar{z}}\mu\para{z,\bar{z}} \sim e^{i\phi_{R}}+\alpha' e^{-i\para{k-1}\phi_{R}} +...\cr
&&\psi_{1,L}\sim \sigma\para{z,\bar{z}}\mu^{\dag}\para{z,\bar{z}} \sim e^{i\phi_{L}}+\alpha' e^{-i\para{k-1}\phi_{L}}+...~, 
\end{eqnarray}
where $...$ denotes higher order (less relevant) terms. The above relations must be viewed as the ``transmutation" of the UV primary fields with a free boson description to the primary fields of the IR fixed point under RG flow. In other words, if we start from the vertex operator on the right side in the UV limit, then RG flow will transform it into the left side as we approach the IR fixed point. We must stress that the lattice parafermion operator that is the building block of some statistical models is different from the parafermion primary field and these two should not be confused. Lattice parafermion does not have a definite chirality, is not a primary field and has a more complicated UV-IR transmutation (see Ref. \onlinecite{clock-3} for more details).  

For some specific values of $k$, the SDSG model can be solved exactly~\cite{Lech-1,Teo-Kane}. In the following we will first describe the exactly solvable $k=2$ SDSG and then discuss the more interesting $k=3$ case. 

\subsection{k=2 Self-dual sine-Gordon model}

At $k=2$, it is possible to fermionize the SDSG model. This can be seen from the scaling dimension of the $\cos\para{2\theta}$ and $\cos\para{2\varphi}$ operator which is unity. These mass terms can be written as a fermion bilinear since the scaling dimension of holomorphic part of free electrons is 1/2. So we are dealing with a non-interacting problem that can be easily solved~\cite{MF-1,Lech-1}. To this end we define the following fermionic operators

\begin{eqnarray} 
 c_{R}\propto e^{i\phi_R}=e^{i\para{\varphi+\theta}},~~ c_{L}\propto e^{i\phi_L}=e^{i\para{\varphi-\theta}},~~ k=2.~~
\end{eqnarray}
Using the above definitions we can rewrite $\mathcal{H}_1$ as
\begin{eqnarray}
\mathcal{H}_1=-1/2\int dx~g_1&&\para{e^{i\para{\phi_R+\phi_L}}+e^{-i\para{\phi_R+\phi_L}}}\cr
+g_2&&\para{e^{i\para{\phi_R-\phi_L}}+e^{-i\para{\phi_R-\phi_L}}} .
\end{eqnarray} 
It is convenient to work with Majorana fermions defined as follows:
\begin{eqnarray}
&&\gamma_{1,R/L}=\frac{c_{R/L}+c_{R/L}^\dag}{2}=\cos\para{\phi_{R/L}},\cr
&&\gamma_{2,R/L}=\frac{c_{R/L}-c_{R/L}^\dag}{2i}=\sin\para{\phi_{R/L}}.
\end{eqnarray}
In the Majorana fermion basis, the sine-Gordon equation for $N=2$ is
\begin{eqnarray}\label{SG N=2}
&& \mathcal{H}_{SG}^{N=2}=\mathcal{H}_0+\mathcal{H}_1,\cr
&&\mathcal{H}_0=i\int dx~\sum_{i=1,2}\para{\gamma_{i,R}\partial_x \gamma_{i,R}-\gamma_{i,L}\partial_x \gamma_{i,L}},\cr
&&\mathcal{H}_1=i\int dx~\para{\para{g_1+g_2} \gamma_{1,R}\gamma_{1,L}-\para{g_1-g_2}\gamma_{2,R}\gamma_{2,L}}.
\end{eqnarray}
where $i$ in the $\mathcal{H}_1$ can be derived by taking the zero mode of Majorana fermions carefully~\cite{Teo-Kane}. At the self-dual point, the interaction Hamiltonian creates $2i g_1  \gamma_{1,R}\gamma_{1,L}$ mass term for the $\gamma_{1,R}$ and $\gamma_{1,L}$ Majorana fermions, and leaves the other two Majorana fermions gapless.  This is consistent with our expectation for $k=2$ SDSG to be described by a $Z_2$ parafermion theory which happens to be the Ising CFT with Majorana fermions as its primary operators.

Now consider an array of coupled 1D chains at $g_1=g_2$ self-dual point. It is clear from Eq. \eqref{SG N=2} that $\gamma_{1,R/L}$ Majorana fermion is gapped, while $\gamma_{2,R/L}\sim \sin\para{\phi_{R/L}}$ remains gapless. We then turn on the coupling between neighboring $\gamma_{2,R/L}$ Majorana fermions. The generic term is of $ig_{I,I+1}\gamma^{I}_{2,R}\gamma^{I+1}_{2,L}$ form, where $I$ stands for the index of the chain. This term gaps out all the Majorana fermions except for the two outermost Majorana modes i.e., $\gamma^{1}_{2,L}$, and $\gamma^{N}_{2,R}$. The bosonized version of inter-chain coupling is as follows:
\begin{eqnarray}
2i\gamma^{I}_{2,R}\gamma^{I+1}_{2,L}=&&2\sin\para{\phi^{I}_{R}}\sin\para{\phi^{I+1}_{L}}\cr
=&&\cos\para{\phi^{I}_{R}+\phi^{I+1}_{L}}-\cos\para{\phi^{I}_{R}-\phi^{I+1}_{L}}.
\end{eqnarray} 

It is worth mentioning that $i\gamma^{I}_{1,R}\gamma^{I+1}_{1,L}=\cos\para{\phi^{I}_{R}}\cos\para{\phi^{I+1}_{L}}$ term can be added to the above mass term without causing a phase transition. As we mentioned above $k=2/\nu_c$, if we start from $\nu_c=1$ quantum Hall state we obtain $k=2$. The above mapping tells us that inducing pairing in a $\nu_c=1$ integer quantum Hall (IQH) state will give us a topological superconductor with Ising anyons. This result can also be shown by starting form a lattice model with $C=1$ Chern number and adding pairing term. The model can be solved exactly after which, it can be shown that there is a range of parameters in which we obtain a topological superconductor with a single Majorana edge state~\cite{Fu-Kane,QHZ}.

\subsection{k=3 Self-dual sine-Gordon model}

The SDSG model at $k=3$ can be shown to flow toward a $Z_3$ parafermion CFT at its IR fixed point~\cite{int-1,int-2,int-3,Lech-1,Lech-2}. One way to show this is to start from a $Z_4$ parafermion theory that can be bosonized. Then by adding an appropriate mass term of the form $:\cos\para{3\theta}:+:\cos\para{3\varphi}:$, it can be shown that the theory will flow to $\mathcal{M}_5$ minimal model which is a
$Z_2$ orbifold of the $Z_3$ parafermion CFT~\cite{Lech-1}. Interestingly, the primary fields of the $Z_4$ parafermion theory transmute under this UV-IR flow as well. Therefore, we can use this UV-IR transmutation to represent the primary fields of the $Z_3$ parafermion in terms of the primaries of the $Z_4$ theory with free boson representation. In Ref.~[\onlinecite{clock-3}], we obtain all the primary fields of the $Z_3$ parafermion theory in terms of current and vertex operators. For example, the following identifications will be shown
\begin{eqnarray}
&&\cos\para{3\phi_{R}/2}\cos\para{3\phi_{L}/2}\sim  X\para{z,\bar{z}},\cr
&& \sin\para{3\phi_{R}/2}\sin{\para{3\phi_{L}/2}} \sim \epsilon\para{z,\bar{z}},
\end{eqnarray}

where $\phi_{R/L}=\varphi\pm \theta$ are chiral bosons, $X\para{z,\bar{z}}$ is the neutral operator and $\epsilon\para{z,\bar{z}}$ is the thermal (energy) operator of the $Z_3$ parafermion theory with $(h_{X},\bar{h}_{X})=(7/5,7/5)$, and $(h_{\epsilon},\bar{h}_{\epsilon})=(2/5,2/5)$ conformal dimensions, respectively~\cite{DF}. Adding neutral operator is an irrelevant perturbation for the resulting IR fixed point CFT. The above relations are consistent with the facts that 1) Both thermal and neutral primary fields are neutral under $Z_3$ and $Z_3^{\rm dual}$ symmetries introduced in Eq. \eqref{Sym-1}. 2) The thermal operator is odd under duality transformation, $\varphi \leftrightarrow \theta$, and neutral operator even. 3) The $\cos\para{3\phi_{R}/2}\cos\para{3\phi_{L}/2}$ perturbation drives the phase transition between $Z_4$ and $Z_3$ theory. 4) Adding $\sin\para{3\phi_{R}/2}\sin{\para{3\phi_{L}/2}}$ term to the $N=3$ SDSG model spoils the self-duality condition and results in a gapped theory. The thermal operator has zero spin and can be written as the product of its holomorphic and anti-holomorphic parts. The above relations suggest the following identifications:
\begin{eqnarray}\label{Thermal-1}
&& \sin\para{3\phi_{R}/2}\sim \epsilon_{R},\quad \sin{\para{3\phi_{L}/2}} \sim \epsilon_{L}.
\end{eqnarray}
It is worth mentioning that $\epsilon^{I}_{R}$ and $ \epsilon^{I+1}_{L}$ operators are not local with respect to themselves since their conformal spins are fractional that amounts to branch-cut in their self-correlation functions. However, $ \epsilon^{I}_{R} \epsilon^{I+1}_{L}$ has zero conformal spin and there is no branch cut in its correlation with itself. So it generates a local perturbation. Recall that chiral and anit-chiral electron operators were bosonized as 

\begin{equation}\label{eq:electron-3}
c_{\pm,R/L}= e^{i\frac{3\phi_{cR/L}}{2}}e^{\pm i\frac{\phi_{sR/L}}{\nu_s}} \quad k=3.
\end{equation}
in the UV limit and at $\nu_c=2/3$ filling (recall that in our notation $c_{+}$ denotes the electron operator on the top layer and $c_{-}$ on the bottom layer). 
Now let us consider the following inter-chain electron pairing and backscattering term. Combining equations \eqref{Thermal-1} and \eqref{eq:electron-3} yields the following transmutation:
\begin{equation}\label{int-Z3-1}
-g_{I,I+1}\Re \para{c^{I\dagger}_{- R}-c^{I}_{+ R}}\para{c^{I+1}_{- L}-c^{I+1\dagger}_{+ L}} \propto -g^*_{I,I+1}\epsilon^{I}_{R}\epsilon^{I+1}_{L},\end{equation}
where $g^*_{I,I+1}=g_{I,I+1}\left\langle\cos\para{\phi^I_{s,R}/\nu_s-\phi^{I+1}_{s,L}/\nu_{s}}\right\rangle$.  Therefore, the inter-chain electron pairing and backscattering can be added to the Hamiltonian of $Z_3$ parafermions to couple neighboring chains in the above fashion. It is known that for $Z_3$ parafermion CFT, perturbing the Hamiltonian with thermal operator moves the theory off the critical point for all values of $g^*_{I,I+1}$~\cite{clock-2}. However, $g^*_{I,I+1}>0$ drives the system into paramagnetic phase, while $ g^*_{I,I+1}<0$ to the ferromagnetic phase.

Consequently, if we add the above inter-chain mass term to the array of gapless $Z_3$ parafermions chains we can gap out all the gapless modes except for the two outermost ones. Next, we appeal to the bulk-boundary conjecture to relate the topological order of the bulk to its edge CFT. Sine the resulting FTSC and Read-Rezayi $Z_3$ parafermion state of the $\nu_c=13/5$ FQH are both gapped in their bulks and have identical non-Abelian edge CFT, therefore, they are described by the same non-Abelian topological order and support the same non-Abelian excitations. Among these excitations, we the most significant one is the Fibonacci anyons that is capable of performing topological quantum computation via braiding operations. 

Now let us imagine an FTSC whose parent state is at filling fraction $\nu_c=2/3$. The value of $k$ in the sine-Gordon model is $k=2/\nu_c=3$. Hence, we obtain the $Z_3$ parafermion FTSC by inducing superconductivity into $\nu_c=2/3$ FQH state. Two fermionic candidates for such a state are $(3,3,0)$, $(2,2,1)$ and $(1,1,2)$ Halperin states. These bilayer FQH states have neutral and charged degrees of freedom each described by $c=1$ CFT. Superconductivity cannot do anything with the neutral part and only the charged part undergoes the phase transition into $Z_3$ parafermion state. However, we must gap out these neutral sector first in order to have excitation with multiples of $q=2/3$ electric charge. Thus, $\nu_c=2/3$ FQH-SC heterostructure can yield the superconducting analogue of the $Z_3$ parafermion FQH state.

\section{Parafermion fractional topological superconductors}\label{sec:four}

\begin{figure}[t]
\includegraphics[width=.5 \textwidth]{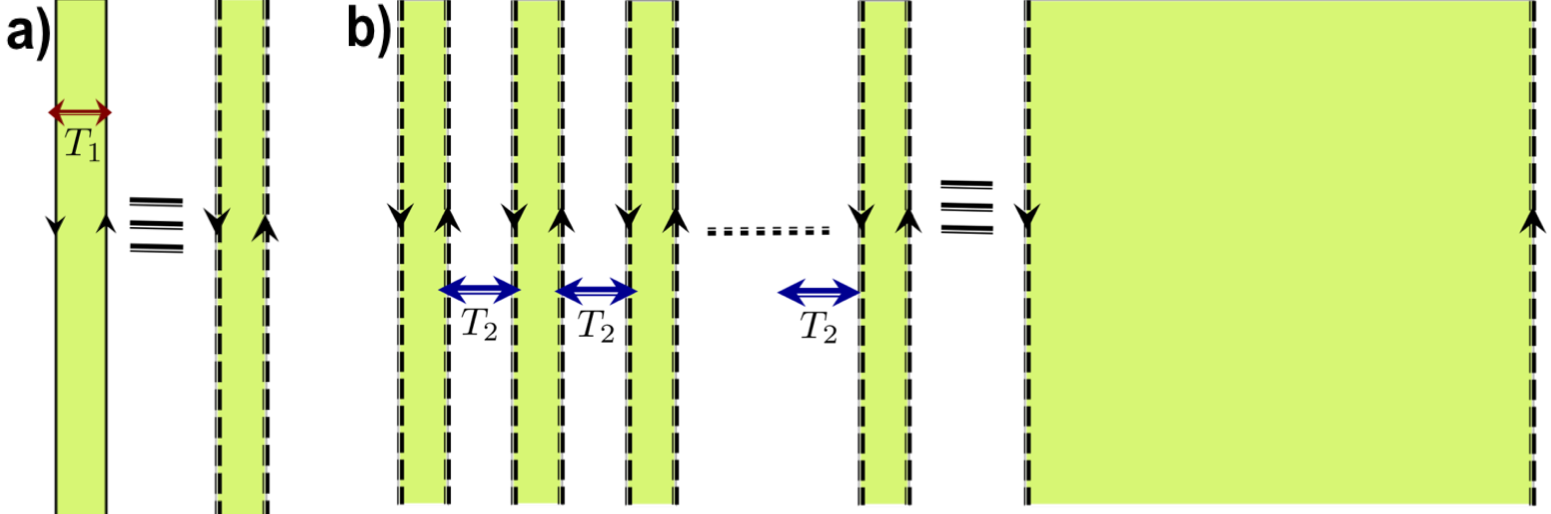}
\caption{Array of coupled 1D chain. a) $T_1$ represents the intra-chain interaction that turns a chain into a parafermion chain. The solid (dashed) line denotes a free boson (parafermion) gapless mode. b) $T_2$ represents the inter-chain coupling that gaps out every two adjacent parafermion modes. After applying $T_1$ and $T_2$ terms, we are left with a gaped bulk and a chiral gapless parafermion edge mode. 
} \label{fig3}
\end{figure}

In this section, we apply the Teo and Kane construction~\cite{Teo-Kane} to obtain a 2D dimensional topological state from an array of 1D gapless chains. To this end, we couple the 1D parafermion chains to one another (see Fig. \ref{fig3}). We first map the 2D FCI at $\nu_c=2/k$ filling fraction into an array of 1D chains of free bosons. When spin-orbit coupling is strong in the parent state we can generate intra-chain electron backscattering through proximity to a ferro-magnetic. On the other hand, the superconducting substrate generates intra-chain electron pairing. Therefore, each chain is described by the SDSG model in Eq. \eqref{SG-2}. Let us call the intra-chain perturbations coming from these two processes $T_{1}^{I,I}$. At the self-dual point i.e., when $g_1=g_2$, the perturbations drive a phase transition from the UV fixed point with $c_{UV}=1$ into the IR fixed point that is now described by $Z_{k}$ parafermion theory with $c_{IR}={2(k-1)}/{(k+2)}$ for $k<4$. Tuning the interaction to the self-dual point is a stringent condition. Deviation from this condition will be addressed shortly. Then we study the effect of intra-chain anyon and electron backscattering along with electron pairing. Let us call this inter-chain perturbation $T_{2}^{I,I+1}$. In section \ref{sec:three} we checked that for $k=2,3$ cases, these terms can be rewritten in terms of the primary fields of the emergent $Z_k$ CFT.  We saw that they correspond to coupling the holomorphic operators on the $I$-th chain to the anti-holomorphic operator on the $(I+1)$-th chain. Such perturbations gap out both gapless modes that are involved. Therefore, by applying $T_{2}^{I,I+1}$ terms we can gap out all the parafermion modes in the system except for the two outermost ones: the left-moving parafermion mode at the first chain and the right-moving parafermion mode at the last chain. Altogether, we obtain a gapped bulk that has a chiral parafermion edge state. Bulk-boundary CFT correspondence implies that all states with the same edge CFT have the same operator content and identical topological orders. Therefore, the state we found is the superconducting analogue of the (non-Abelian part of the) Read-Rezay parafermion states originally proposed for the FQH systems.

Now let us discuss the effect of moving away from the self-dual point of the SDSG model in Eq. \eqref{SG-2}. Since the 2D $Z_k$ parafermion state that we obtained by starting from the self-dual point ($g_1=g_2$) is gapped, it is robust against perturbations and maintains its topological properties as long as perturbations do not close the gap. The bulk gap is determined by the inter-chain interaction $T_{2}^{I,I+1}$. Assuming, it opens a mass gap, $\Delta$, we expect the system to stay in the $Z_k$ parafermion state as long as the distance from the self-dual point which is proportional to $\abs{g_1-g_2}$ does not exceed $\Delta$. Therefore, the exact self-duality condition is not necessary.

We would to emphasize again that the above picture applies to $k=2/\nu_c< 4$ cases only, i.e., when $\nu_c \in \{1, 2/3\}$. For other values of $k$, the cosine terms in Eq. \eqref{SG-2} are irrelevant and do not modify the low energy physics drastically. However, our simple RG analysis may fail for other cases and we may obtain $Z_k$ parafermion FTSC in other cases as well. One reason is that our RG analysis is based on 1D CFT calculation, while the system that we have in hand is in fact a 2D system.  If we believe in the RG analysis based on the 1D CFT, the low energy physics (IR limit) of the SDSG is described by a $c=1$ CFT for $k> 4$. We conjecture that this theory is a $U(1)_k/Z_2$ orbifold CFT. The unperturbed Hamiltonian of the 1D chains in Eq. \eqref{SG-2}(i.e., when $g_1=g_2=0$) is described by a $U(1)_k$ CFT of free bosons. In this case, the chiral algebra is generated by the current operator, $j=\partial \phi$, and creation $c_{R/L}^{\dag}$ and annihilation $c_{R/L}$ operators, where $c_{R/L}\sim e^{ik\phi_{R/L}}$. However, superconductivity breaks the $U(1)$ symmetry associated with the charge degree of freedom. So we cannot have both $c_{R/L}$ and $c_{R/L}^\dag$ in the chiral algebra. One possibility is that perturbation makes $i\para{c_{R/L}-c_{R/L}^\dag}\sim \sin\para{k\phi_{R/L}}$ operator gapped while maintains the other combination $c_{R/L}+c_{R/L}^\dag\sim \cos\para{k\phi_{R/L}}$ gapless. In this case, the chiral algebra is generated by current operator $j$ and $\cos\para{k\phi_{R/L}}$ operators only. The latter operator is invariant under the following discrete $Z_2$ symmetry: $\phi_{R/L}\to -\phi_{R/L}$. Therefore, we can identify $\phi_{R/L}$ with $-\phi_{R/L}$. The resulting chiral algebra is well-studied and gives rise to the $U(1)_k/Z_2$ orbifold CFT. Interestingly, this theory supports twist operators with quantum dimensions equal to $d_{\rm tw}=\sqrt{k}$ which are physical manifestations of the bulk superconducting vortices, namely electron operator picks a negative sign when it circles around twist operators~\cite{FTSC-1}. Therefore, we conjecture that an FTSC whose parent state is at $\nu_c=2/k<1/2$ filling fraction is described by a $U(1)_k/Z_2$ orbifold CFT instead of a $Z_k$ parafermion theory.  

\section{Summary and conclusion}\label{sec:five}
In this paper, we explained that a fractional quantum Hall system at $\nu_c=2/k$ filling fraction can be mapped to an array of narrow strips each with a pair of counter-propagating gapless modes. These narrow strips can be viewed as quasi-1D chains whose charged sectors are described by a $U(1)$ CFT. The coupling between the right moving branch of the $I$-th chain and the left moving branch of the $(I+1)$-th chain gaps out the bulk but maintains the edge modes of the 2D system gapless. If we assume spin-orbit coupling is so large in the system that spin of electron is locked to its momentum and as a result all right (left) moving sectors carry spin up (down), then we can induce s-wave paring as well as in-plane magnetic order in the bulk of the system via proximity or intrinsic spontaneous symmetry breaking.  

By inducing superconductivity in the bulk of a fractional quantum Hall we can obtain other possibilities besides the above mentioned mechanism for achieving the Abelian phase from coupled 1D chains. As we discussed in the current paper, we can first couple the right and left moving parts of each chain (intra-chain interaction) and tune the strength of pairing and backscattering to obtain gapless $Z_{k}$ parafermion chains. Then, adjacent parafermion modes can be coupled to gap out the bulk of the system. The edge of the system is gapless, chiral, and is described by a $Z_{k}$ parafermion CFT. We can understand this result in a more intuitive way. Th electron operator in the $\nu_c=2/k$ FQH state is equivalent to $k/2$ anyons (when all neutral fluctuations are gapped). Accordingly, a Cooper pair consists of $k$ anyons. As a result, we expect a topological state with $k$-anyon condensation to emerge through the proximity effect. One candidate for such a condensate is the Read-Rezayi $Z_{k}$ parafermion state. The most interesting case is a $\nu_c=2/3$ FQH on top of a superconducting substrate. By increasing the strength of pairing the system undergoes a topological phase transition into the $Z_3$ parafermion FTSC. The operator content of this state contains Fibbonacci anyons that are capable of performing universal quantum computation. The $\nu_c=2/3$ FQH can be achieved in $(3,3,0)$, $(2,2,1)$ or $(1,1,2)$ bilayer states. For $(3,3,0)$ bilayer state, we need to add an interlayer pairing otherwise the two layers would act independently and would have $\nu_{\uparrow}=\nu_{\downarrow}=1/3$ effective fillings. 

\section{Acknowledgement}

We gratefully acknowledge useful discussion with E.-A. Kim, T. Senthil, M.-Z. Hasan, N. Regnault, S. Raghu, and S.-B. Chung. We are in particular thankful to Maissam Barkeshli for his insightful comments and valuable discussions. This work was supported by NSF CAREER grant DMR-0955822.

{\noindent} {\em Note added. ---} After the completion of this work, we became aware of a recently posted work~[\onlinecite{Mong-2013}] with a related but different topological order for $2/3$ FQH-SC hetero-structure setup.


\begin{thebibliography}{63}
\bibitem{comment-1}
It should be emphasized again that in the case of multi-component (multi-layer) FQH states, $e^{i\phi_{R/L}/\nu_c}$ only represents the charged part of the electron operator. The neutral part is described by other bosonic degrees of freedom that are not affected by superconducting pairing. We assume all neutral sectors are gapped.
\end{thebibliography}
\end{document}